\documentclass[11pt]{article}

\usepackage{graphicx}
\usepackage{caption,subcaption}
\usepackage{gnuplot-lua-tikz}

\usepackage[colorlinks=true,linkcolor=red,urlcolor=blue]{hyperref}

\usepackage{cite}

\usepackage{setspace}
\onehalfspace

\begin{document}
\begin{center}
{\huge Identification of discontinuities\\in plasma plume evolution\footnote{The original manuscript is accepted for publication by Applied Physics B: Lasers and Optics, and it is available at www.springerlink.com, \href{http://link.springer.com/article/10.1007/s00340-013-5440-3}{doi: 10.1007/s00340-013-5440-3}.}}

\vspace{20pt}
{\sc Ardian B. Gojani\footnote{Institute of Fluid Science, Tohoku University, gojani@edge.ifs.tohoku.ac.jp. Author acknowledges the support of BK21 program at Seoul National University and GCOE program at Tohoku University.} \& Jahja Kokaj\footnote{Department of Physics, Kuwait University.} \& Shigeru Obayashi\footnote{Institute of Fluid Science, Tohoku University.}}

\vspace{20pt}
\end{center}

\begin{spacing}{1}
\noindent \textbf{Abstract:} The ejection of material during laser ablation gives rise to the development of discontinuities in the ambient gas. Several of these discontinuities are observed and characterized, including externally and internally propagating shock waves, contact surface, and the ionization front. Qualitative experimental observations and analysis of these discontinuities is presented. Results from shadowgraphy enabled determination of an irradiance threshold between two different ablation mechanisms, and determination of several stages of plasma plume evolution. Consideration of the refractive index as a dynamic sum of the contributions from gas and electrons led to separate identification of ionization front from the contact surface. Furthermore, ionization front was observed to lead the shock wave at the earlier stage of the ablation.
\end{spacing}

\section{Introduction}
Laser ablation is the term used for the process of removal of a small amount of material from its surface (usually, of the order of a few tens of ng to a $\mu$g) by means of a focused laser beam. Results of extensive studies of laser ablation can be found in the textbook authored by B\"{a}uerle \cite{Bau_Laser_11}, while a wide range of applications are discussed in the books \cite{CH_Pulsed_94,Phi_Laser_07,Eas_Pulsed_07}. Due to complexities of the processes during ablation, the exact description of ablation mechanisms at different irradiations for different materials remains a difficult problem. This problem is usually attacked by determining threshold points in the transition of the behavior of the ablation as a function of irradiance (or, equivalently, fluence). The range between two threshold points would define a single dominant mechanism for ablation, which is then modeled by analytical or computational methods. A complication arises when ablation involves plasma that interacts with the incoming laser beam \cite{MH_lpi_10}. 

From the experimental point of view, ablation mechanisms are deduced from the results obtained by several techniques, which can be divided into two general types: techniques that investigate the material surface after ablation, and those that investigate the dynamics of the ablation plume. The outcome of the first type is the volume of the ablated material or the ablation rate, which can be measured by confocal microscopy surface profiles \cite{GY_nae_ASS09}, or the morphology of the ablated area, which can be determined by scanning electron microscopy \cite{Goj_EFM12}. The second type of experiments includes visualization and spectroscopy techniques, which measure the shape and velocity of the ablation plume, density and temperature of the plume, and variation of plume composition, all these resolved in space and time. An example of the application of these latter techniques is the study of the ablation process for lithotripsy \cite{KMM_sml_FA03, KM_icu_OLT04, KM_lsi_OLE07, KM_oid_OLE08}. The interest in plasma generation and shock propagation in laser lithotripsy is that the mechanisms of stone destruction and removal are not yet fully understood, in particular the removal of ablated material from the bulk of the stone and the coupling of opto-mechanical properties for efficient stone destruction.

Ablation is invariably coupled with the rapid ejection of material, which leads to formation of several dynamic zones in the surrounding ambient. This paper presents a study of these zones, with emphasis on the observation of discontinuities and their evolution. Since the approach is qualitative, only a simple analytical model is taken into consideration, supported by experimental results. Shadowgraphy is a well suited and one of the most accessible techniques for visualization of these discontinuities. In particular, the rapid ejection of the plasma plume and the subsequent evolution of discontinuities in the ambient gas easily can be observed with relatively good spatial and temporal resolution, while densities and temperatures can be assessed only qualitatively. 

The number of publications on laser ablation that rely on shadowgraphy is enormous, but as an example of the use of the technique might serve the well known papers by Callies \emph{et al.} \cite{CBH_tro_JPDAP995} and Brietling \emph{et al.} \cite{BS5_sii_APA999}. Similarly to these references, our paper reports the observation of discontinuities due to ablation, namely externally and internally propagating shock waves, contact surface, and ionization front. But, unlike most published works, our paper draws attention to the observation that, in a certain stage of the evolution of the plasma plume, the ionization front leads the shock wave.

\section{Refractivity of plasma}
\begin{figure*}
\centering{
\begin{minipage}{\textwidth}
\centering{\input{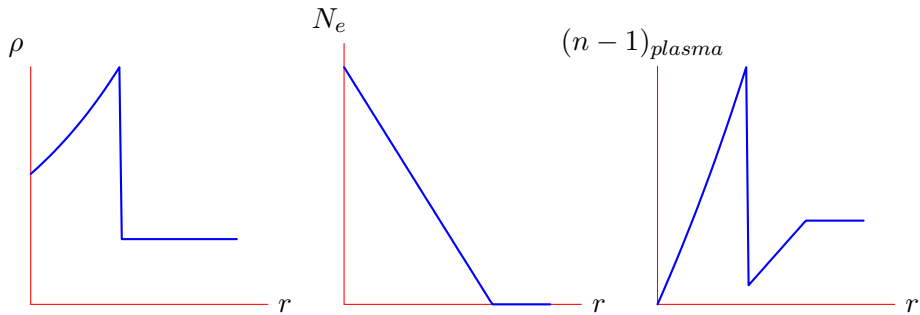}}
\end{minipage}
\caption{Qualitative illustration of the change of refractive index.}\label{figindex}
}
\end{figure*}

For the ablation laser energies used in this study (irradiation range $I = 6 - 130$ GW/cm$^2$), material removal always is coupled with the creation of the plasma. In other words, the ablated plume consists of free electrons, ions, and neutral atoms. Since the process is dynamic, the number densities of these constituents change with time, and at later stages (order of $\mu$s), even clusters and congregated microparticles are formed \cite{YJ5_epe_APL00}. In addition, during the entire ablation process, the plume interacts with the surrounding air, causing the change of its density, mainly through the effect of the shock waves \cite{WM4_ela_JAP07}.

As it is well known, shadowgraphy registers the displacement of light due to gradients in the refractive index, yielding images with zones of varying brightness. The inhomogeneities in the medium through which light is propagating cause the light to bend, and the angle of deflection $\varepsilon$ is related to the refractive index $n(\textbf{r})$ by equation \cite{BW_Optics_05}
\begin{equation}\label{angle}
\hat{\textbf{u}}\, d\varepsilon = \nabla \ln n(\textbf{r})\, ds,
\end{equation}
where $ds$ is the infinitesimal path that light traverses and $\hat{\textbf{u}}$ is the unit normal vector. Refractive index of plasma is a combination of the refractive indices of its constituents, i.e. neutral gas, ions, and electrons \cite{JS_orp_971}. As it is discussed by Hermoch \cite{Her_smi_CJPB970}, in a first approximation of the dispersion theory, the refractive indices of a gas consisting only of neutrals and a gas consisting only of ions, are considered to be of the same order of magnitude. The overall contribution of neutral and ionic gases on the refractive index of the plasma is linearly proportional to the concentration of constituents, i.e. density. Therefore, their combined refractivity is expressed by the Gladstone-Dale relation
\begin{equation}\label{ngas}
\left( n-1 \right)_{\rho} \propto \rho,
\end{equation}
which holds for a wide range of densities. 

On the other hand, the refractive index for the electron gas is 
\begin{equation}\label{nel}
n_e^2 = 1 - \frac{\omega_p^2}{\omega^2} =1 - \frac{N_e}{N_c},
\end{equation}
where $\omega_p$ is the plasma frequency, $\omega = c/\lambda_v$ is the frequency of the visualization light, $N_e$ is the electron number density, and 
\begin{equation}\label{ncutoff}
N_c = \frac{\omega^2 m_e \epsilon_0}{e^2}
\end{equation}
is the cutoff density ($c$ speed of light, $m_e$ electron mass, $\epsilon_0$ vacuum permittivity, $e$ electron charge). Equation (\ref{nel}) shows that when \begin{equation}\label{nenc}
N_e > N_c,
\end{equation}
$n_e$ is imaginary, and as a consequence, the visualization light is reflected by plasma, and only the near field evanescent wave manages to pass through. As the electron number density decreases, plasma transmittance increases and equation (\ref{nel}) can be approximated by relation
\begin{equation}\label{nelectron}
\left( n-1 \right)_{e} \propto - N_e.
\end{equation}

The overall refractive index of the plasma, then, can be expressed as
\begin{equation}\label{nplasma}
\left( n-1 \right)_{plasma} = \left( n-1 \right)_{\rho} + \left( n-1 \right)_{e}. 
\end{equation}
Obviously, this equation is not an exact one, but it can be useful in demonstrating the trend of the change of the refractivity of plasma. In this simple case, the two-gas model is applied to plasma and influences from density gradient and electron number density variation can be determined.

Substitution of equations (\ref{ngas}) and (\ref{nelectron}) into (\ref{nplasma}) shows that refractivity of plasma is a contribution of two opposing effects on the angle $\hat{\textbf{u}} \varepsilon$, as expressed by (\ref{angle}). Local increase of gas density $\rho$ has the effect that it increases $n$ locally, producing a gradient and bending the light in the direction of higher density. This is the familiar case of observation of shock waves, which are recorded as adjoined narrow dark and bright lines in shadowgraph images. The effect of electrons, though, is opposite: the increase of electron number density reduces $n$, thus bending the light away from the zone with high $N_e$.

The effect of the main discontinuities on the refractivity of plasma is illustrated in figure {\ref{figindex}, where density jump due to shock wave is modeled on a simple blast wave, and where a linear decrease of electron number density with distance $r$ from the ablation center is assumed (a study of several profile functions for electron number density can be found in \cite{Kei_isi_PP972}). The effect on the refractive index is the generation of two discontinuities with gradients of opposite sign, which can be recorded by shadowgraphy, as demonstrated by the following experiment.

\section{Experiment}
The experiment was done by ablating metallic targets (aluminum and copper) in air using a laser beam with $\lambda_a = 1064$ nm, $\tau_a = 7$ ns and $E_a = 17 - 360$ mJ, focused onto a spot with diameter $d = 200\, \mu$m. The choice of metallic targets for this study is obvious, because they readily provide an excess of free electrons. Placement of the sample was chosen such that air breakdown, judged by the luminous volume of the plasma in the absence of the target, would be a few millimeters behind the position designated for target placement.

The visualization was done by a standard shadowgraph system, consisting of a laser beam with $\lambda_v = 532$ nm and $\tau_v = 3$ ns used for illumination, and an ICCD image sensor ($1280 \times 1024$ pixels, pixel size $6.7 \times 6.7$ $\mu$m$^2$) for image recording. The visualization beam was expanded and collimated by a diverging-converging lens combination, and directed parallel to the material surface, perpendicular to the ablating laser beam. A single converging lens was used for imaging. The expanded beam had a diameter about two times larger than the diagonal of the image sensor.

Visualization presents two challenges: since plasma emits a very intense light, a light source that overpowers plasma's self-luminosity is required. On the other hand, the camera sensor should not be overexposed. Our approach was to use a combination of a laser beam as a powerful light source with a series of neutral density filters as adjusters of illumination that reaches the image sensor. Optical density of ND filters was $\approx 2.7$ (transmittance $\approx 0.2\%$), and they were placed just in front of the camera, so that the entire plasma and ambient light is blocked. Using laser as a light source also provided the benefit of having extremely short exposures. The visualization laser light had a pulse energy of 25 mJ, but since it was expanded to a large area, its influence on the plasma was negligible. 

Scattered light from the exit port of the ablation laser was collected by a high-speed biased photodetector (rise time 1 ns) and used as the triggering signal for the entire system. Triggering was controlled by a pulsed delay generator and was applied to the visualization laser and the camera.

\section{Results}

\begin{figure*}[!t]
\centering
\begin{subfigure}{.3\textwidth}
\includegraphics[width=\textwidth]{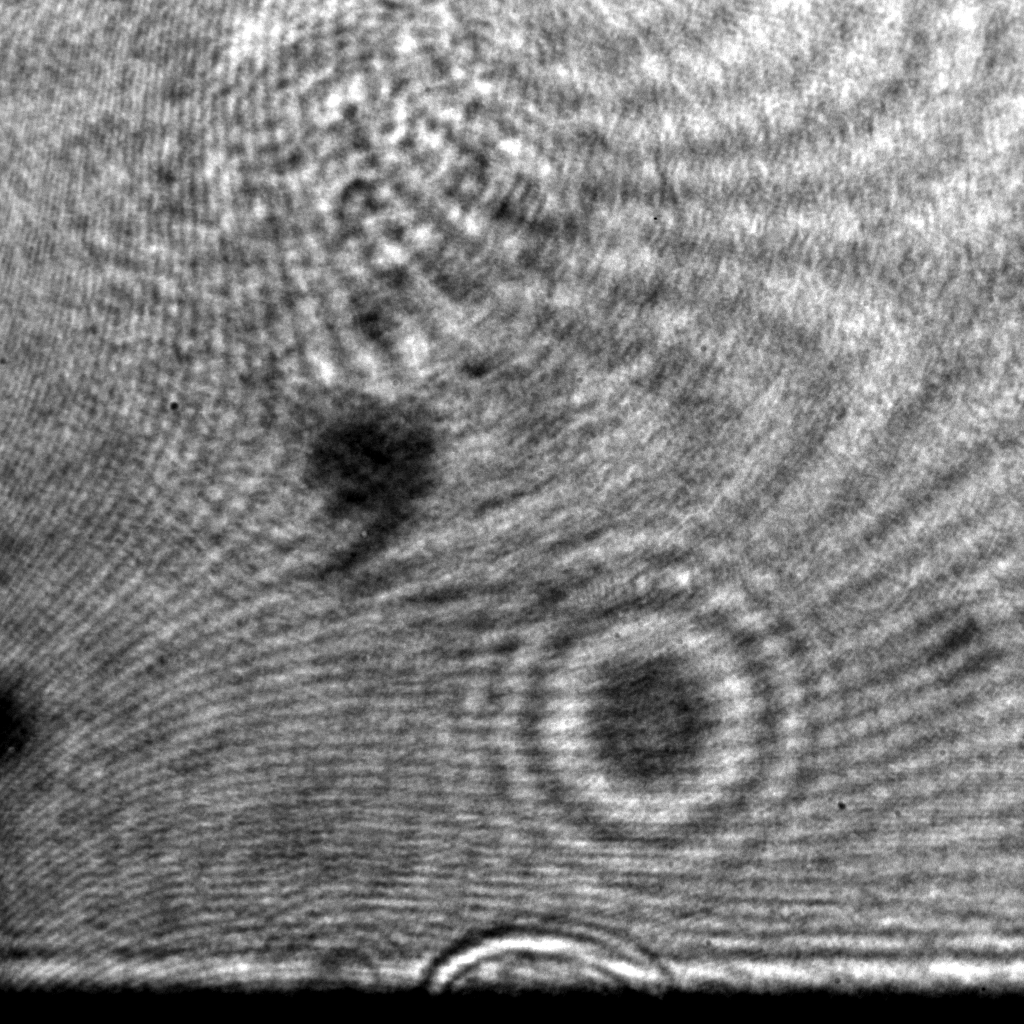}
\caption{$\Delta t = 40$ ns}\label{fig17a}
\end{subfigure}
\begin{subfigure}{.3\textwidth}
\includegraphics[width=\textwidth]{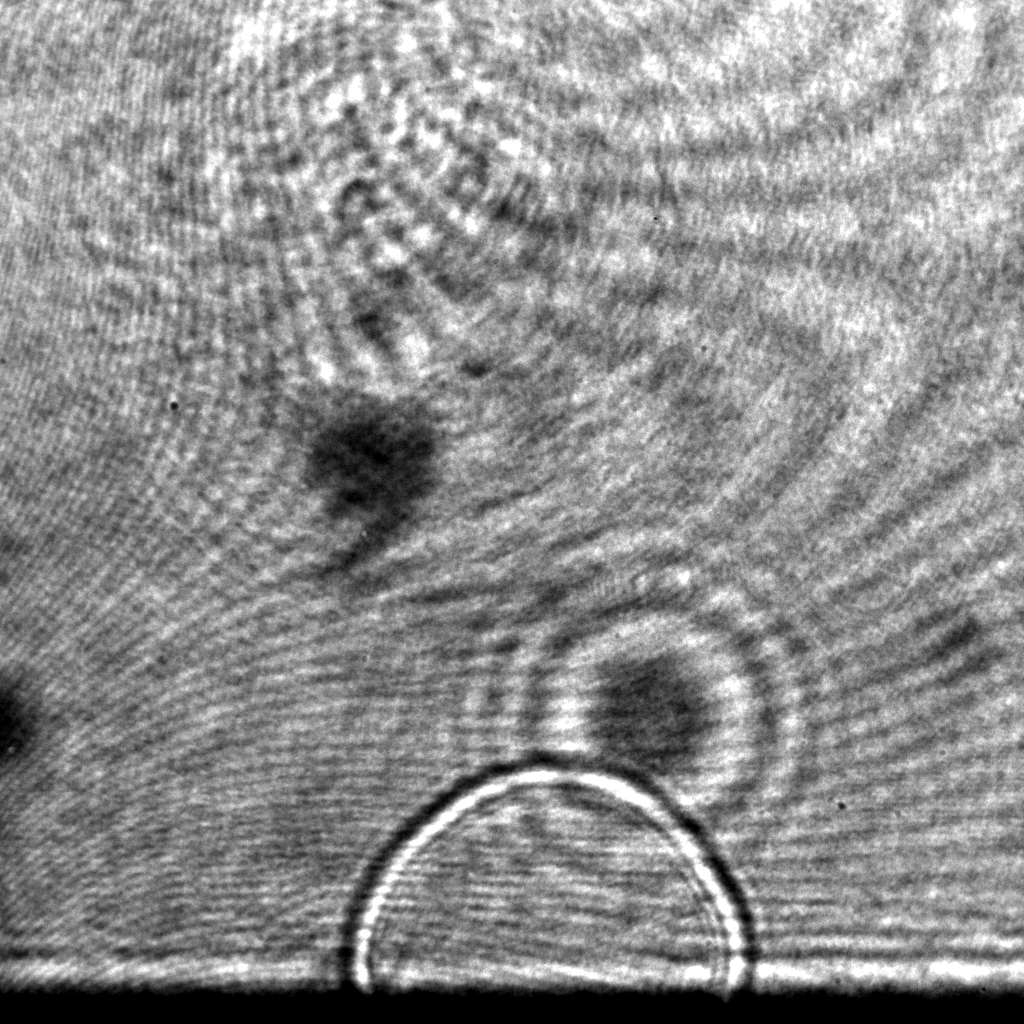}
\caption{$\Delta t = 180$ ns}\label{fig17b}
\end{subfigure}
\caption{Discontinuities from ablation of aluminum by 17 mJ laser energy.}\label{fig17mJ}
\end{figure*}

\begin{figure*}[!t]
\centering
\begin{subfigure}{.3\textwidth}
\includegraphics[width=\textwidth]{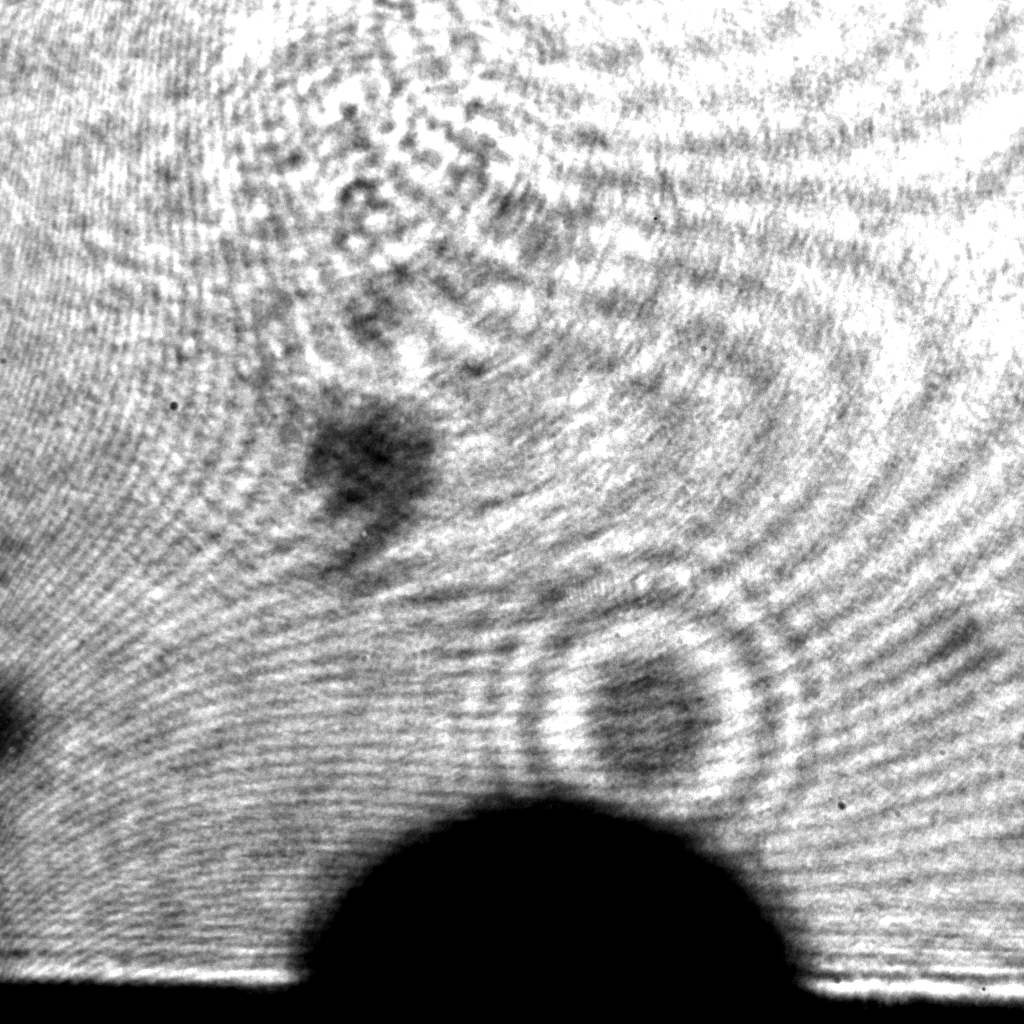}
\caption{$\Delta t = 40$ ns}\label{fig25a}
\end{subfigure}
\begin{subfigure}{.3\textwidth}
\includegraphics[width=\textwidth]{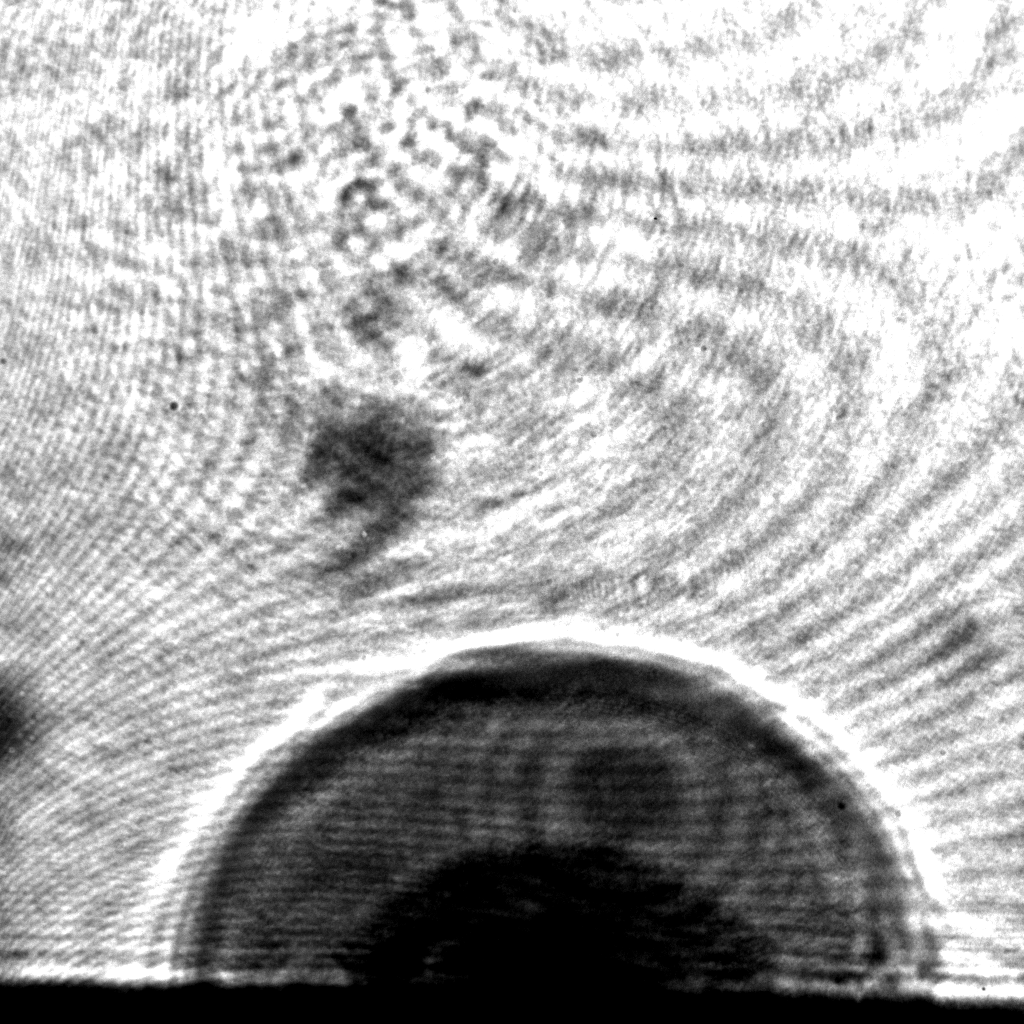}
\caption{$\Delta t = 80$ ns}\label{fig25b}
\end{subfigure}
\begin{subfigure}{.3\textwidth}
\includegraphics[width=\textwidth]{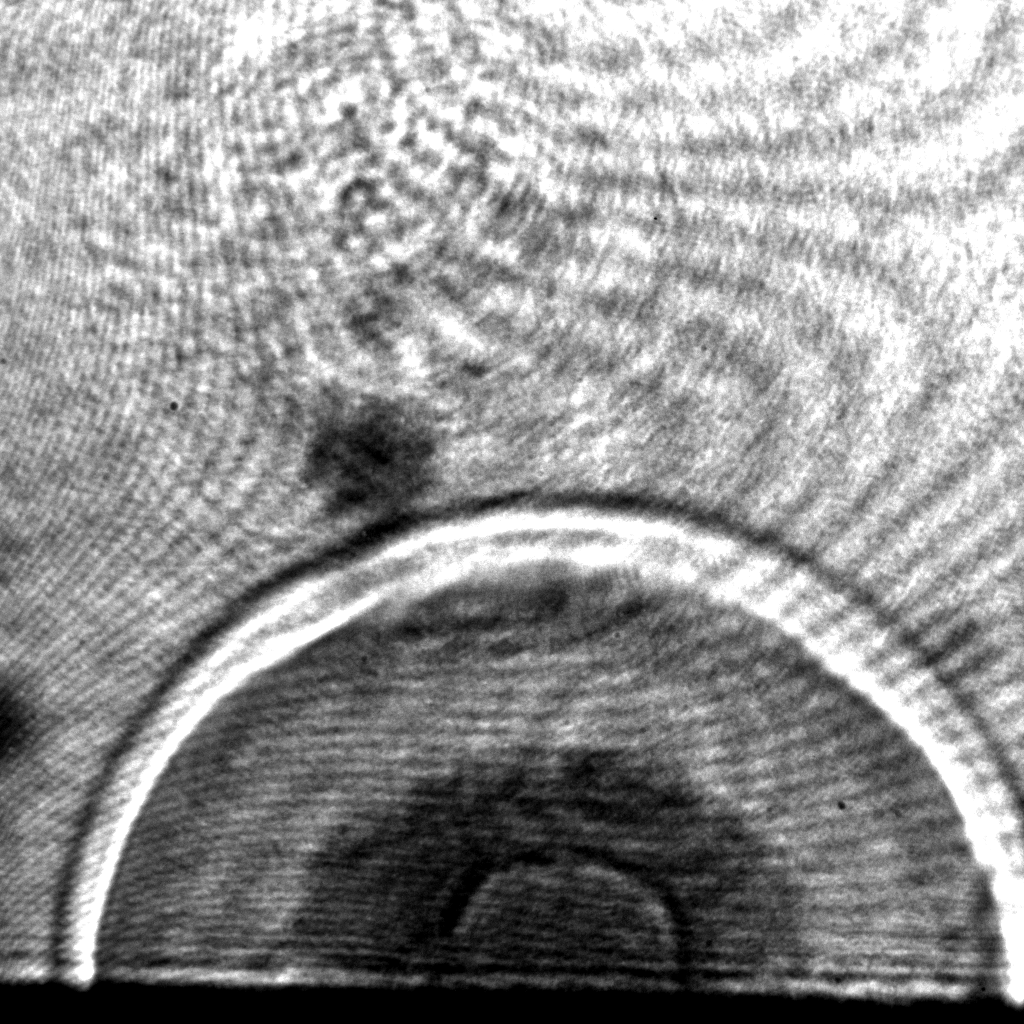}
\caption{$\Delta t = 140$ ns}\label{fig25c}
\end{subfigure}
\caption{Same as in figure \ref{fig17mJ}, but now laser energy is 25 mJ.}\label{fig25mJ}
\end{figure*}

Two examples of visualization results are presented in figures \ref{fig17mJ} and \ref{fig25mJ}, which show the discontinuities due to ablation of aluminum by irradiances of $I=6$ GW/cm$^2$ and $I=10$ GW/cm$^2$, respectively. An image frame corresponds to the field of view of $2.1 \times 2.1$ mm$^2$ (2.1 $\mu$m per pixel), and the time when the image is taken is noted in the caption. Time $\Delta t$ is defined as the time that has passed from the moment when the first motion of the material is observed to the moment when the pulse of the visualization laser is fired. It should be noted that the initial time of the ejection of the material depends slightly on the ablation laser energies, which might come about due to different mechanisms that lead to ablation: evaporation and normal boiling, or phase explosion \cite{KM_mtm_NIMPRB997, Lu_tep_PRE03}.

Figures reveal the disadvantage of using a laser beam for visualization: numerous interference fringes that arise from the neutral density filters and protective windows attached to the image sensor produce images of average quality. Nevertheless, discontinuities are clearly observed. Since the visualization laser can be triggered to emit only a single pulse, each image corresponds to a fresh ablation. Comparison of images obtained under the same conditions reveal that the ablation was done with excellent repeatability. Images for ablation laser energies higher than 25 mJ show the same features as those of figure \ref{fig25mJ}. Similar results are obtained for copper as well.

Figure \ref{fig17mJ} shows the familiar sight of a shock wave in the form of a blast, expanding away from the ablation zone. The visualization of the shock shows a leading dark stripe followed by the bright stripe. No plasma plume is observed in this series of figures. Figure \ref{fig25mJ}, on the other hand, is much more complex. Initially, as seen in figure \ref{fig25a}, visualization of ablation is characterized only by an opaque hemispherical plasma plume. At a later stage, (figure \ref{fig25b}) some morphology of the plasma plume can be observed. The visualization of ablation is led by a white stripe of light, followed by a dark one. The transmittance of the plasma plume inside these stripes varies, with a semitransparent and an opaque zone clearly identifiable. At a much later stage, as shown in figure \ref{fig25c}, the leading stripe is dark, immediately followed by a bright stripe. This is the familiar effect of a shock wave in shadowgraphy. Following is a band of bright stripe, and the interior of the plasma plume now shows even more complex morphology: plasma transmittance has increased, and plasma plume and internally propagating shock waves have become visible. At much later times, the visualization becomes qualitatively identical to that of figure \ref{fig17mJ}, with only the shock wave propagating away from the ablation zone being visible, until the shock is dissipated into a sound wave.

\section{Discussion}
The dramatic difference between the discontinuities visualized in figures \ref{fig17mJ} and \ref{fig25mJ} is an indication that the ablation mechanisms were different, but shadowgraphy is very limited in offering an explanation. Still, based on discussions from Section 2, several observations can be made.

According to equation (\ref{ncutoff}), the cutoff electron number density for our experiment is $N_c \approx 10^{20}$ cm$^{-3}$. For the ablation shown in figure \ref{fig17mJ}, this number is not achieved, because plasma is transparent throughout the process and only the shock wave is observed.

The next level of irradiance tested, achieves the cutoff value close to the target and for initial moments of plume ejection, as it can be inferred from figures \ref{fig25a} and \ref{fig25b}, where plasma is opaque. Further evidence for the large electron number density is provided by figure \ref{fig25b}. Based on equation \ref{nplasma}, the refractivity of plasma is dominated by the contribution of electrons, because the light is being deflected away from the target. The effect of electrons is observable even at later stages (as in figure \ref{fig25c}), although now shock waves have expanded and precede. Since the lower limit of electron number density that would effect the refractivity of plasma is of the order of $10^{14}$ cm$^{-3}$, it can be concluded that $N_e$ is greater than this value. Order-of-magnitude comparison to spectroscopic measurements further support this conclusion \cite{Goj_esl_ISRN12}.

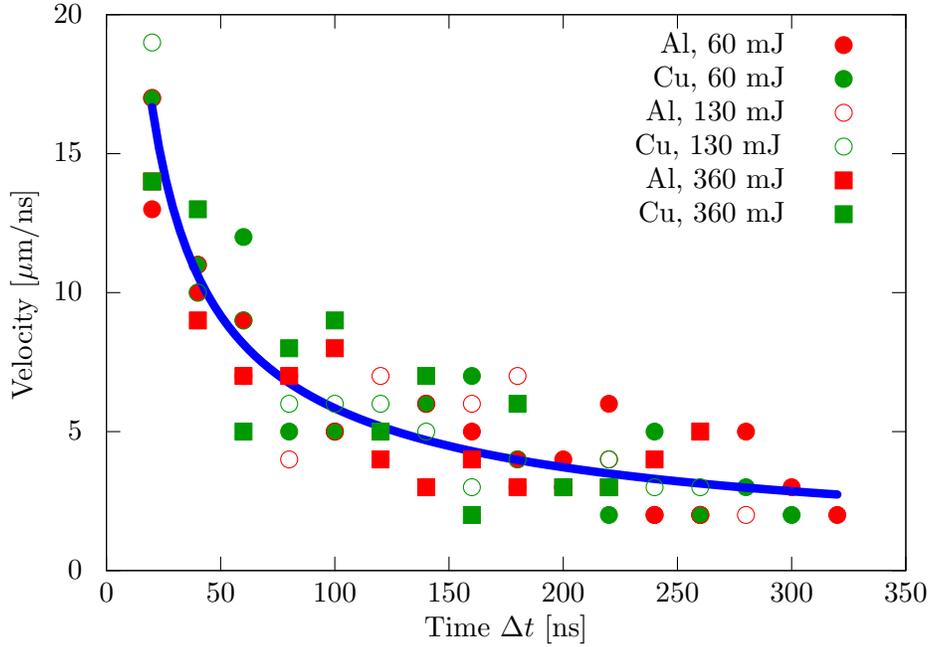
\begin{figure}[!ht]
\centering{
\begin{minipage}{\textwidth}
\centering{\begin{tikzpicture}[gnuplot]
\gpsolidlines
\gpcolor{gp lt color border}
\gpsetlinetype{gp lt border}
\gpsetlinewidth{1.00}
\draw[gp path] (1.320,0.985)--(1.500,0.985);
\draw[gp path] (11.947,0.985)--(11.767,0.985);
\node[gp node right] at (1.136,0.985) { 0};
\draw[gp path] (1.320,2.834)--(1.500,2.834);
\draw[gp path] (11.947,2.834)--(11.767,2.834);
\node[gp node right] at (1.136,2.834) { 5};
\draw[gp path] (1.320,4.683)--(1.500,4.683);
\draw[gp path] (11.947,4.683)--(11.767,4.683);
\node[gp node right] at (1.136,4.683) { 10};
\draw[gp path] (1.320,6.532)--(1.500,6.532);
\draw[gp path] (11.947,6.532)--(11.767,6.532);
\node[gp node right] at (1.136,6.532) { 15};
\draw[gp path] (1.320,8.381)--(1.500,8.381);
\draw[gp path] (11.947,8.381)--(11.767,8.381);
\node[gp node right] at (1.136,8.381) { 20};
\draw[gp path] (1.320,0.985)--(1.320,1.165);
\draw[gp path] (1.320,8.381)--(1.320,8.201);
\node[gp node center] at (1.320,0.677) { 0};
\draw[gp path] (2.838,0.985)--(2.838,1.165);
\draw[gp path] (2.838,8.381)--(2.838,8.201);
\node[gp node center] at (2.838,0.677) { 50};
\draw[gp path] (4.356,0.985)--(4.356,1.165);
\draw[gp path] (4.356,8.381)--(4.356,8.201);
\node[gp node center] at (4.356,0.677) { 100};
\draw[gp path] (5.874,0.985)--(5.874,1.165);
\draw[gp path] (5.874,8.381)--(5.874,8.201);
\node[gp node center] at (5.874,0.677) { 150};
\draw[gp path] (7.393,0.985)--(7.393,1.165);
\draw[gp path] (7.393,8.381)--(7.393,8.201);
\node[gp node center] at (7.393,0.677) { 200};
\draw[gp path] (8.911,0.985)--(8.911,1.165);
\draw[gp path] (8.911,8.381)--(8.911,8.201);
\node[gp node center] at (8.911,0.677) { 250};
\draw[gp path] (10.429,0.985)--(10.429,1.165);
\draw[gp path] (10.429,8.381)--(10.429,8.201);
\node[gp node center] at (10.429,0.677) { 300};
\draw[gp path] (11.947,0.985)--(11.947,1.165);
\draw[gp path] (11.947,8.381)--(11.947,8.201);
\node[gp node center] at (11.947,0.677) { 350};
\draw[gp path] (1.320,8.381)--(1.320,0.985)--(11.947,0.985)--(11.947,8.381)--cycle;
\node[gp node center,rotate=-270] at (0.246,4.683) {Velocity [$\mu$m/ns]};
\node[gp node center] at (6.633,0.215) {Time $\Delta t$ [ns]};
\node[gp node right] at (10.479,7.976) {Al, 60 mJ};
\gpcolor{gp lt color 0}
\gpsetpointsize{8.00}
\gppoint{gp mark 7}{(1.927,5.792)}
\gppoint{gp mark 7}{(2.535,4.683)}
\gppoint{gp mark 7}{(3.142,4.313)}
\gppoint{gp mark 7}{(3.749,3.574)}
\gppoint{gp mark 7}{(4.356,2.834)}
\gppoint{gp mark 7}{(4.964,2.464)}
\gppoint{gp mark 7}{(5.571,3.204)}
\gppoint{gp mark 7}{(6.178,2.834)}
\gppoint{gp mark 7}{(6.785,2.464)}
\gppoint{gp mark 7}{(7.393,2.464)}
\gppoint{gp mark 7}{(8.000,3.204)}
\gppoint{gp mark 7}{(8.607,1.725)}
\gppoint{gp mark 7}{(9.214,1.725)}
\gppoint{gp mark 7}{(9.822,2.834)}
\gppoint{gp mark 7}{(10.429,2.094)}
\gppoint{gp mark 7}{(11.036,1.725)}
\gppoint{gp mark 7}{(11.121,7.976)}
\gpcolor{gp lt color border}
\node[gp node right] at (10.479,7.526) {Cu, 60 mJ};
\gpcolor{gp lt color 1}
\gppoint{gp mark 7}{(1.927,7.272)}
\gppoint{gp mark 7}{(2.535,5.053)}
\gppoint{gp mark 7}{(3.142,5.423)}
\gppoint{gp mark 7}{(3.749,2.834)}
\gppoint{gp mark 7}{(4.356,2.834)}
\gppoint{gp mark 7}{(4.964,2.834)}
\gppoint{gp mark 7}{(5.571,3.204)}
\gppoint{gp mark 7}{(6.178,3.574)}
\gppoint{gp mark 7}{(6.785,2.094)}
\gppoint{gp mark 7}{(7.393,2.094)}
\gppoint{gp mark 7}{(8.000,1.725)}
\gppoint{gp mark 7}{(8.607,2.834)}
\gppoint{gp mark 7}{(9.214,1.725)}
\gppoint{gp mark 7}{(9.822,2.094)}
\gppoint{gp mark 7}{(10.429,1.725)}
\gppoint{gp mark 7}{(11.121,7.526)}
\gpcolor{gp lt color border}
\node[gp node right] at (10.479,7.076) {Al, 130 mJ};
\gpcolor{gp lt color 0}
\gppoint{gp mark 6}{(1.927,7.272)}
\gppoint{gp mark 6}{(2.535,5.053)}
\gppoint{gp mark 6}{(3.142,3.574)}
\gppoint{gp mark 6}{(3.749,2.464)}
\gppoint{gp mark 6}{(4.356,2.834)}
\gppoint{gp mark 6}{(4.964,3.574)}
\gppoint{gp mark 6}{(5.571,3.204)}
\gppoint{gp mark 6}{(6.178,3.204)}
\gppoint{gp mark 6}{(6.785,3.574)}
\gppoint{gp mark 6}{(7.393,2.094)}
\gppoint{gp mark 6}{(8.000,2.464)}
\gppoint{gp mark 6}{(8.607,1.725)}
\gppoint{gp mark 6}{(9.214,1.725)}
\gppoint{gp mark 6}{(9.822,1.725)}
\gppoint{gp mark 6}{(11.121,7.076)}
\gpsetlinetype{gp lt plot 0}
\gpcolor{gp lt color 2}
\gpsetlinewidth{8.00}
\draw[gp path] (1.927,7.150)--(2.019,6.609)--(2.111,6.173)--(2.203,5.814)--(2.295,5.512)%
  --(2.387,5.254)--(2.479,5.030)--(2.571,4.833)--(2.663,4.659)--(2.755,4.504)--(2.847,4.364)%
  --(2.939,4.238)--(3.031,4.123)--(3.123,4.018)--(3.215,3.921)--(3.307,3.831)--(3.399,3.749)%
  --(3.491,3.672)--(3.583,3.600)--(3.675,3.533)--(3.767,3.470)--(3.859,3.411)--(3.951,3.355)%
  --(4.043,3.303)--(4.135,3.253)--(4.227,3.206)--(4.319,3.162)--(4.411,3.119)--(4.503,3.079)%
  --(4.596,3.040)--(4.688,3.003)--(4.780,2.968)--(4.872,2.935)--(4.964,2.902)--(5.056,2.871)%
  --(5.148,2.842)--(5.240,2.813)--(5.332,2.786)--(5.424,2.759)--(5.516,2.734)--(5.608,2.709)%
  --(5.700,2.686)--(5.792,2.663)--(5.884,2.641)--(5.976,2.619)--(6.068,2.598)--(6.160,2.578)%
  --(6.252,2.559)--(6.344,2.540)--(6.436,2.522)--(6.528,2.504)--(6.620,2.487)--(6.712,2.470)%
  --(6.804,2.454)--(6.896,2.438)--(6.988,2.422)--(7.080,2.407)--(7.172,2.393)--(7.264,2.379)%
  --(7.356,2.365)--(7.448,2.351)--(7.540,2.338)--(7.632,2.325)--(7.724,2.313)--(7.816,2.300)%
  --(7.908,2.288)--(8.000,2.276)--(8.092,2.265)--(8.184,2.254)--(8.276,2.243)--(8.368,2.232)%
  --(8.460,2.222)--(8.552,2.211)--(8.644,2.201)--(8.736,2.191)--(8.828,2.182)--(8.920,2.172)%
  --(9.012,2.163)--(9.104,2.154)--(9.196,2.145)--(9.288,2.136)--(9.380,2.128)--(9.472,2.119)%
  --(9.564,2.111)--(9.656,2.103)--(9.748,2.095)--(9.840,2.087)--(9.932,2.079)--(10.024,2.072)%
  --(10.116,2.064)--(10.208,2.057)--(10.300,2.050)--(10.392,2.043)--(10.484,2.036)--(10.576,2.029)%
  --(10.668,2.022)--(10.760,2.016)--(10.852,2.009)--(10.944,2.003)--(11.036,1.997);
\gpcolor{gp lt color border}
\node[gp node right] at (10.479,6.626) {Cu, 130 mJ};
\gpcolor{gp lt color 1}
\gpsetlinewidth{1.00}
\gppoint{gp mark 6}{(1.927,8.011)}
\gppoint{gp mark 6}{(2.535,4.683)}
\gppoint{gp mark 6}{(3.142,4.313)}
\gppoint{gp mark 6}{(3.749,3.204)}
\gppoint{gp mark 6}{(4.356,3.204)}
\gppoint{gp mark 6}{(4.964,3.204)}
\gppoint{gp mark 6}{(5.571,2.834)}
\gppoint{gp mark 6}{(6.178,2.094)}
\gppoint{gp mark 6}{(6.785,2.464)}
\gppoint{gp mark 6}{(7.393,2.094)}
\gppoint{gp mark 6}{(8.000,2.464)}
\gppoint{gp mark 6}{(8.607,2.094)}
\gppoint{gp mark 6}{(9.214,2.094)}
\gppoint{gp mark 6}{(11.121,6.626)}
\gpcolor{gp lt color border}
\node[gp node right] at (10.479,6.176) {Al, 360 mJ};
\gpcolor{gp lt color 0}
\gppoint{gp mark 5}{(1.927,6.162)}
\gppoint{gp mark 5}{(2.535,4.313)}
\gppoint{gp mark 5}{(3.142,3.574)}
\gppoint{gp mark 5}{(3.749,3.574)}
\gppoint{gp mark 5}{(4.356,3.943)}
\gppoint{gp mark 5}{(4.964,2.464)}
\gppoint{gp mark 5}{(5.571,2.094)}
\gppoint{gp mark 5}{(6.178,2.464)}
\gppoint{gp mark 5}{(6.785,2.094)}
\gppoint{gp mark 5}{(7.393,2.094)}
\gppoint{gp mark 5}{(8.000,2.094)}
\gppoint{gp mark 5}{(8.607,2.464)}
\gppoint{gp mark 5}{(9.214,2.834)}
\gppoint{gp mark 5}{(11.121,6.176)}
\gpcolor{gp lt color border}
\node[gp node right] at (10.479,5.726) {Cu, 360 mJ};
\gpcolor{gp lt color 1}
\gppoint{gp mark 5}{(1.927,6.162)}
\gppoint{gp mark 5}{(2.535,5.792)}
\gppoint{gp mark 5}{(3.142,2.834)}
\gppoint{gp mark 5}{(3.749,3.943)}
\gppoint{gp mark 5}{(4.356,4.313)}
\gppoint{gp mark 5}{(4.964,2.834)}
\gppoint{gp mark 5}{(5.571,3.574)}
\gppoint{gp mark 5}{(6.178,1.725)}
\gppoint{gp mark 5}{(6.785,3.204)}
\gppoint{gp mark 5}{(7.393,2.094)}
\gppoint{gp mark 5}{(8.000,2.094)}
\gppoint{gp mark 5}{(11.121,5.726)}
\gpcolor{gp lt color border}
\gpsetlinetype{gp lt border}
\draw[gp path] (1.320,8.381)--(1.320,0.985)--(11.947,0.985)--(11.947,8.381)--cycle;
\gpdefrectangularnode{gp plot 1}{\pgfpoint{1.320cm}{0.985cm}}{\pgfpoint{11.947cm}{8.381cm}}
\end{tikzpicture}
}
\end{minipage}
\caption{External shock front velocity as a function of time.}\label{shpejtesia}}
\end{figure}

Ablation studies by visualization make use of the blast wave theory \cite{Sed_Similarity_993}, where the pressure and temperatures involved in the ablation process are calculated from the measurements of shock radius \cite{PW_trn_JPDAP09}. The evolution of the shock radius is described by equation
\begin{equation}\label{eqradius}
r(t)=C \left(\frac{E_{esw}}{\rho_0}\right)^{1/(2+\xi)} t^{2/(2+\xi)},
\end{equation}
where $C$ is a constant (usually $C \approx 1$), $E_{esw}$ is the energy content of the shock wave, and $\rho_0$ is the density of the surrounding atmosphere. $\xi$ is a parameter that accounts for the dimensionality of the propagation, and it is equal to 3 for spherical propagation of the shock wave\footnote{In the previous version, this value was reported the incorrect value 2. This correction does not affect the subsequent discussion and conclusions, because the analysis of the results was done with $\xi=3$. In particular, curve fitting shown in figure \ref{shpejtesia} corresponds to $v(t) \propto t^{-0.65}$. Authors thank Krste Pangovski of the Department of Engineering, Cambridge University, for pointing at the error.}. Spherical symmetry of the propagation of the shock wave is an evidence that plasma being initiated by the breakdown of the solid, rather than the breakdown of the surrounding air, because, in the latter case, plasma would appear elongated along the Rayleigh range, exhibiting a cylindrical symmetry.

Results for shock velocities as a function of time are shown in figure \ref{shpejtesia}. Points show the velocity calculated as the difference of the radius of the shock between two successive images, divided by the time interval between those images. The curve is the derivative of $r(t)$ from equation (\ref{eqradius}). Since the energy content of the shock wave does not change much with the increase of irradiance, a single velocity curve is shown for illustrative purposes.

Applying deduced shock velocity to strong shock relations, we can easily calculate the pressure and temperature behind the shock wave \cite{ZR_Shock_966}. The relation between pressure and shock velocity is given by 
\begin{equation}
P = \frac{\rho_0}{\gamma+1} \left(\frac{dr}{dt}\right)^2,
\end{equation}
where $\gamma$ is specific heat ratio of the surrounding ambient (for air, $\gamma=7/5$), while the relationship for temperature is
\begin{equation}
T = T_0\frac{\gamma-1}{\gamma+1} \frac{P}{P_0},
\end{equation}
with $T_0$ and $P_0$ being room temperature and pressure, respectively. Calculations show that within the first 300 ns, pressure and temperature of the shocked gas just behind the external shock wave decrease from 300 MPa and 100 000 K down to 0.5 MPa and 2 500 K. Since the shock Mach number throughout this time interval is larger then 10, it can be stated that the shock wave maintains the plasma by partially ionizing the ambient air.

\begin{figure}[!t]
\centering
\begin{minipage}{0.8\textwidth}
\center{\input{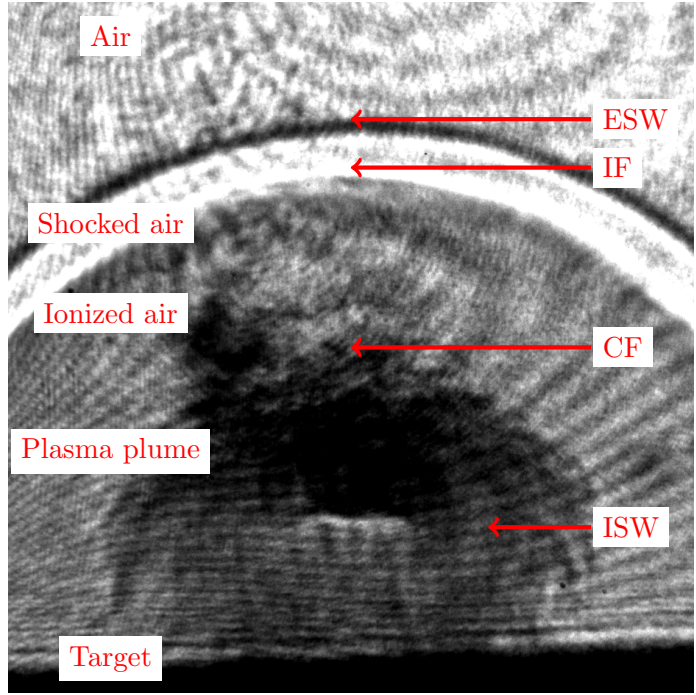}}
\end{minipage}
\caption{Ablation discontinuities and zones. Ablation laser energy used 92 mJ. $\Delta t = 180$ ns.}\label{figdiscont}
\end{figure}

In summary, different discontinuities of the ablated plasma plume can be defined as in figure \ref{figdiscont}: ESW - externally propagating shock wave, IF - ionization front, CF - contact front, ISW - internally propagating shock wave. ESW divides the laboratory air and the shocked air, which is then followed by the zone dominated by an excess of electrons - the ionized air. Since initially the ionization front was observed to be ahead of the shock wave, but later passed by the shock, the ionized air is initiated by the electrons from the material, but later, the strong shock wave is able to maintain a level of ionization. Contact front is the surface that designates the beginning of the ablated material - plasma plume. Its constituents are cooled ions and neutrals from the target that collide with the air molecules. Since the target is continually ejecting hot material, this leads to the creation of the internal shock wave, which initially moves away, but later moves towards the target. Based on figure \ref{fig25mJ}, a rough estimate of time stages for visualization can be set as follows: during the first stage ($\approx 50 - 60$ ns) the number density of electrons is larger than the cutoff value, therefore the plasma is opaque, because it completely reflects the visualization beam. During the second stage ($\approx 60 - 80$ ns), the ionization front leads the shock wave, and plasma is semi transparent. During the third stage ($\approx 80-160$ ns), external shock wave overtakes the ionization front, internal shock wave dissipates, and the entire ablation process can be modelled as a blast wave. Finally, at later stages, only the external shock wave is visualized.

\section{Conclusions}
In this paper, the observation and interpretation of discontinuities following laser ablation were discussed based on the change of plasma refractivity. It was shown that shadowgraphy provides a simple, but useful method for identifying laser ablation mechanisms and for plasma diagnostics, although only order-of-magnitude quantitative results can be obtained. Specifically, a transition in the ablation mechanisms was observed between irradiances 6 and 10 GW/cm$^2$. The approximate electron number density was judged based on the transmittance of the plasma.

A more useful result from shadowgraphy was the ability to identify several discontinuities in the ambient surrounding the ablation spot, and measure their evolution with good spatial and temporal resolution. This allowed for characterization of the ablation plume and definition of zones of different constituents, based on their effect on the refractive index. Special attention was drawn to the observation of ionization front leading the externally propagating shock wave, which indicated that initially plasma was generated by ejected electrons from the target, and only at later stages the shock wave was able to induce air ionization.

\begin{spacing}{1}

\end{spacing}


\begin{thebibliography}{99}

\bibitem{Bau_Laser_11}
D.~B\"{a}uerle. {\em Laser Processing and Chemistry}. Springer, Berlin, 2011.

\bibitem{CH_Pulsed_94}
D.~B. Chrisey and G.~K.~Hubler (eds.). {\em Pulsed Laser Deposition of Thin Films}. Wiley, New York, 1994.

\bibitem{Phi_Laser_07}
C.~Phipps (ed.). {\em Laser Ablation and its Applications}. Springer, New York, 2007.

\bibitem{Eas_Pulsed_07}
R.~Eason (ed.). {\em Pulsed Laser Deposition of Thin Films}. Wiley, New Jersey, 2007.

\bibitem{MH_lpi_10}
I.~N. Mihailescu and J.~Hermann. Laser-plasma interactions. In P. Schaaf (ed.) {\em Laser Processing of Materials: Fundamentals, Applications and Developments}, pages 49--88. Springer, Berlin, 2010

\bibitem{GY_nae_ASS09}
A.~B. Gojani and J.~J. Yoh. New ablation experiment aimed at metal expulsion at the hydrodynamic regime. {\em Appl Surf Sci}, 255:9268–--9272, 2009.

\bibitem{Goj_EFM12}
A.~B. Gojani. Laser ablation at hydrodynamic regime. {\em Eur Phys J Conf}, EFM12:217--223, 2013.

\bibitem{KMM_sml_FA03}
J.~Kokaj, M.~Marafi, and J.~Mathews. Study of the mechanisms of laser-based lithotripsy using optical techniques. {\em Fizika A}, 12:151--160, 2003.

\bibitem{KM_icu_OLT04}
J.~Kokaj and J.~Mathews. Imaging and correlation used for guided laser lithotripsy. {\em Opt Laser Technol}, 36:441--448, 2004.

\bibitem{KM_lsi_OLE07}
M.~Marafi, J.~Kokaj, K.~S. Bahtia, Y.~Makdisi, and J.~Mathews. Laser spectroscopy and imaging of gallbladder stones, tissue and bile. {\em Opt Laser Eng}, 45:191--197, 2007.

\bibitem{KM_oid_OLE08}
J.~Kokaj, M.~Marafi, and J.~Mathews. Optical investigation of dynamics of phenomena of laser-based lithotripsy. {\em Opt Laser Eng}, 46:535--540, 2008.

\bibitem{CBH_tro_JPDAP995}
G.~Callies, P.~Berger, and H.~H\"{u}gel. Time-resolved observation of gas-dynamic discontinuities arising during excimer laser ablation and their interpretation. {\em J Phys D: Appl Phys}, 28:794 -- 806, 1995.

\bibitem{BS5_sii_APA999}
D.~Brietling, H.~Schittenhelm, P.~Berger, F.~Dausinger, and H.~H\"{u}gel. Shadowgraphic and interferometric investigation on Nd:YAG laser-induced vapor/plasma plumes for different processing wavelengths. {\em Appl Phys A}, 69:S505--S508, 1999.

\bibitem{YJ5_epe_APL00}
J.~H. Yoo, S.~H. Jeong, X.~L. Mao, R.~Greif, and R.~E. Russo. Evidence for phase-explosion and generation of large particlesduring high power nanosecond laser ablation of silicon. {\em Appl Phys Let}, 76:783--785, 2000.

\bibitem{WM4_ela_JAP07}
S.~B. Wen, X.~Mao, R.~Greif, and R.~E. Russo. Expansion of the laser ablation vapor plume into a background gas. I. Analysis. {\em J Appl Phys}, 101:023114--1--13, 2007.

\bibitem{BW_Optics_05}
M.~Born and E.~Wolf. {\em Principles of Optics}. Cambridge University Press, Cambridge, 2005.

\bibitem{JS_orp_971}
F.~C. Jahoda and G.~A. Sawyer. Optical refractivity of plasmas. In R. H. Lovberg and H. R. Griem (eds.) {\em Methods in Experimental Physics, Vol 9B}, pages 1--48. Academic Press, New York, 1971.

\bibitem{Her_smi_CJPB970}
V.~Hermoch. Schlieren method and its application for plasma diagnostics. {\em Czech J Phys B}, 20:939--949, 1970.

\bibitem{Kei_isi_PP972}
F.~Keilmann. An infrared schlieren interferometer for measuring electron density profiles. {\em Plasma Physics}, 14:111--122, 1972.

\bibitem{KM_mtm_NIMPRB997}
R.~Kelly and A.~Miotello. On the mechanisms of target modification by ion beams and laser pulses. {\em Nucl Instrum Meth B}, 122:374--400, 1997.

\bibitem{Lu_tep_PRE03}
Q.~Lu. Thermodynamic evolution of phase explosion during high-power nanosecond laser ablation. {\em Phys Rev E}, 67:016410--1--5, 2003.

\bibitem{Goj_esl_ISRN12}
A.~B. Gojani. Experimental study of laser-induced brass and copper plasma for spectroscopic applications. {\em ISRN Spectroscopy}, 2012:868561--1--8, 2012.

\bibitem{Sed_Similarity_993}
L.~I. Sedov. {\em Similarity and Dimensional Methods in Mechanics}. CRC Press, Boca Raton, 1993.

\bibitem{PW_trn_JPDAP09}
C.~Porneala and D.~A. Willis. Time-resolved dynamics of nanosecond laser-induced phase explosion. {\em J Phys D: Appl Phys}, 42:155503--1--7, 2009.

\bibitem{ZR_Shock_966}
Ya.~B. Zeldovich and Yu.~P. Raizer. {\em Physics of Shock Waves and High-Temperature Hydrodynamic Phenomena}. Academic Press, London, 1966.


\end{thebibliography}
\end{document}